\documentclass[prd,showpacs,amsmath,amssymb,twocolumn]{revtex4}
\usepackage{graphicx,color}
\begin{document}

\title{The dispersive
contribution of $\rho(1450,1700)$ decays and $X(1576)$}

\author{Bo Zhang}\author{Xiang Liu}\email{xiangliu@pku.edu.cn}
\author{Shi-Lin Zhu}\email{zhusl@phy.pku.edu.cn}
\affiliation{Department of physics, Peking University, Beijing,
100871, China}

\vspace*{1.0cm}

\date{\today}
\begin{abstract}
We study whether the broad enhancement $X(1576)$ arises from the
final state interaction (FSI) of $\rho(1450,1700)\to \rho^+\rho^-\to
K^{+}K^{-}$ decays. We consider both the absorptive and dispersive
contribution of the above amplitudes since the intermediate states
are very close to $\rho(1450,1700)$. The same mechanism leads to a
similar enhancement around 1580 MeV in the $\pi^{+}\pi^-$ spectrum
in the $J/\psi\to \pi^{0}\pi^{+}\pi^{-}$ channel, which can be used
to test whether $X(1576)$ can be ascribed to the FSI effect of
$\rho(1450,1700)\to \rho^+\rho^-$.
\end{abstract}

 \maketitle

\section{introduction}

A broad enhancement $X(1576)$ with $J^{PC}I^G=1^{--}1^+$ was
observed by BES Collaboration in $K^{+}K^{-}$ invariant mass
spectrum in the $J/\psi\to \pi^{0}K^{+}K^{-}$ channel
\cite{1576-BES}. Its resonance parameters are
$m=(1576^{+49}_{-55}(\mathrm{stat})^{+98}_{-91}(\mathrm{syst}))
-i(409^{+11}_{-12}(\mathrm{stat})^{+32}_{-67}(\mathrm{syst}))$
MeV. The branching ratio is $B[J/\psi\to X(1576)\pi^{0}]\cdot
B[X(1576)\to K^{+}K^{-}]=(8.5\pm 0.6^{+2.7}_{-3.6})\times
10^{-4}$. Its extremely large width around 800 MeV motivated
theoretical explanations such as a $K^{*}(892)-\kappa$ molecular
state \cite{Guo-1576}, tetraquark \cite{Lipkin-1576,1576-QSR},
diquark-antidiquark bound state \cite{Ding-1576,zhang}, a
composition of $\rho(1450)$ and $\rho(1700)$ \cite{Li}.

Since there are two broad overlapping resonances $\rho(1450)$ and
$\rho(1700)$ with the same quantum number around 1600 MeV, we
investigated whether such a broad signal could be produced by the
final state interaction (FSI) \cite{FSI} effect in our previous work
\cite{Liu}. We noticed that the interference effect of
$\rho(1450,1700)$ could produce an enhancement around $1540$ MeV
with the opening of the $\rho\rho$ channel, similar to the cusp
effect discussed in Ref. \cite{bugg}. However, the branching ratio
$B[J/\psi\to \pi^{0}\rho(1450,1700)]\cdot B[\rho(1450,1700)\to
K^{+}K^{-}]$ from the FSI effect was far less than the experimental
data. It's important to point out that we considered only the
contribution of the absorptive part in Ref. \cite{Liu}.

Recently Meng and Chao explored the possible assignment of
$X(3872)$ as the $\chi_{c1}^\prime$ candidate \cite{chao}. They
found the dispersive part of the FSI amplitude contributes more
importantly to the hidden charm decay width of $X(3872)$ than the
imaginary part derived in Ref. \cite{plb} because the intermediate
states $D^0{\bar D}^{0\ast}$ and $D^-D^{+\ast}$ lie very close to
$X(3872)$. Motivated by the above observation, we investigate the
potential role of the dispersive contribution of FSI since the
$\rho\rho$ intermediate states are rather close to
$\rho(1450,1700)$.

This paper is organized as follows. The formulation of
$\rho(1450,1700)\to \rho^+\rho^-\to K^+K^-$ by exchanging
$K^{0(*)}$ is presented in Section \ref{sec2} and our numerical
result and discussion in Section \ref{sec3}.

\section{formulation}\label{sec2}

\begin{figure}[htb]
\begin{center}
\begin{tabular}{cc}
\scalebox{0.8}{\includegraphics{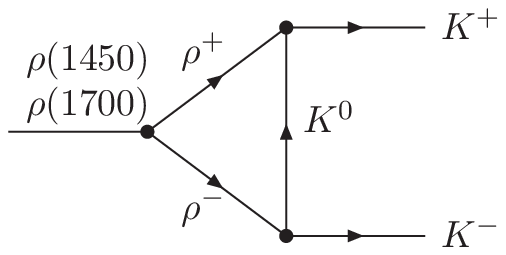}}&\scalebox{0.8}{\includegraphics{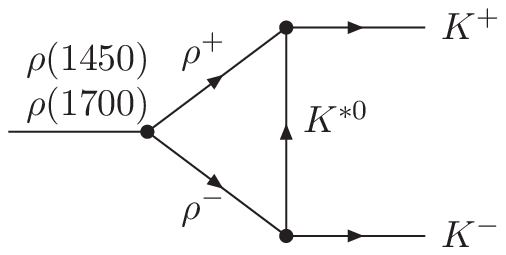}}\\
(a)&(b)\end{tabular}
\end{center}
\caption{The decay of $\rho(1450,1700)\to K^{+}K^{-}$ through the
$\rho^{\pm}$ pair.}\label{KK}
\end{figure}

As shown in Fig. \ref{KK}, we focus on the FSI of the $\rho\rho$
intermediate states through the exchange of the $K^{0(*)}$ meson:
$\rho(1450,1700)\to \rho^+\rho^-\to K^+K^-$.

In order to derive the absorptive amplitude, we introduce the
effective Lagrangians
\begin{eqnarray}
\mathcal{L}_{V_{1}\to P_{1}P_{2}}&=&ig_{1}(P_{1}{\stackrel{\leftrightarrow}{\partial}}P_{2})V^{\nu},\\
\mathcal{L}_{V_{1}\to
V_{2}P_{1}}&=&g_{1}\epsilon_{\mu\nu\alpha\beta}V_{1}^{\mu}\partial^{\nu}P_{1}\partial^{\beta}V_{2}^{\alpha},
\\
\mathcal{L}_{V_{1}\to
V_{2}V_{3}}&=&ig_{2}\Big\{V_{1}^{\mu}(\partial_{\mu}V_{2}^{\nu}V_{3\nu}-V_{2}^{\nu}\partial_{\mu}V_{3\nu})\nonumber\\&&
+(\partial_{\mu}V_{1\nu}V_{2}^{\nu}-V_{1\nu}\partial_{\mu}V_{2})V_{3}^{\mu}\nonumber\\&&+
V_{2}^{\mu}(V_{1}^{\nu}\partial_{\mu}V_{3\nu}-\partial_{\mu}V_{1\nu}V_{3}^{\nu})
\Big\},
\end{eqnarray}
where $g_{i}$'s denote the coupling constants. $P_{i}$ and $V_{i}$
respectively denote the pseudoscalar and vector fields.

\subsection{Absorptive contribution}

Using the Cutkosky cutting rule, one obtains the the absorptive
contribution to the process of
$\rho(1450,1700)\to\rho^{+}(p_{1})\rho^{-}(p_{2})\to
K^{+}(p_{3})K^{-}(p_{4})$ with the exchanged mesons $K^{0}$ and
$K^{*0}$
\begin{eqnarray}
&&\mathbf{Abs}^{(a)}[\rho^{+}\rho^{-},K^{0}]\nonumber\\
&&=\frac{1}{2}\int\frac{d^{3}p_{1}}{(2\pi)^{3}2E_{1}}
\frac{d^{3}p_{2}}{(2\pi)^{3}2E_{2}}(2\pi)^{4}\delta^{4}
(P-p_{1}-p_{2})\nonumber\\&&\times\Big\{ig_{\rho(1450)\rho\rho}[\epsilon\cdot(p_{1}-p_{2})g_{\mu\nu}-\epsilon_{\mu}
(2p_{1}+p_{2})_{\nu}\nonumber\\&&+\epsilon_{\nu}(2p_{2}+p_{1})_{\mu}]\Big\}[ig_{_{\rho
KK}}(q_{\alpha}+p_{3\alpha})]\nonumber\\&&\times[ig_{_{\rho
KK}}(q_{\beta}-p_{4\beta})]\bigg[-g^{\nu\alpha}+\frac{p_{1}^{\nu}p_{1}^{\alpha}}{m_{\rho}^{2}}\bigg]
\nonumber\\&& \times
\bigg[-g^{\mu\beta}+\frac{p_{2}^{\mu}p_{2}^{\beta}}{m_{\rho}^{2}}\bigg]
\bigg[\frac{i}{q^2
-m_{K}^{2}}\bigg]\mathcal{F}^{2}(m_{K},q^2),\label{kk-1}
\end{eqnarray}
and {\small{\begin{eqnarray}
&&\mathbf{Abs}^{(b)}[\rho^{+}\rho^{-},K^{*0}]\nonumber\\
&&=\frac{1}{2}\int\frac{d^{3}p_{1}}{(2\pi)^{3}2E_{1}}
\frac{d^{3}p_{2}}{(2\pi)^{3}2E_{2}}(2\pi)^{4}\delta^{4}
(P-p_{1}-p_{2})\nonumber\\&&\times\Big\{ig_{\rho(1450)\rho\rho}
[\epsilon\cdot(p_{1}-p_{2})g_{\mu\nu}-\epsilon_{\mu}
(2p_{1}+p_{2})_{\nu}\nonumber\\&&+\epsilon_{\nu}(2p_{2}+p_{1})_{\mu}]\Big\}
[ig_{\rho
KK^{*}}\epsilon_{\alpha\beta\kappa\gamma}p_{1}^{\alpha}q^{\kappa}]\nonumber\\&&\times\Big[
ig_{\rho
KK^{*}}\epsilon_{\xi\lambda\delta\zeta}p_{2}^{\xi}q^{\delta}\Big]
\Big[-g^{\nu\beta}+\frac{p_{1}^{\nu}p_{1}^{\beta}}{m_{\rho}^{2}}\Big]
\nonumber\\&& \times
\Big[-g^{\mu\lambda}+\frac{p_{2}^{\mu}p_{2}^{\lambda}}{m_{\rho}^{2}}\Big]
\Big[-g^{\gamma\zeta}+\frac{q^{\gamma}q^{\zeta}}{m_{K^{*}}^{2}}\Big]
\nonumber\\&& \times\Big[\frac{i}{q^2
-m_{K^{*}}^{2}}\Big]\mathcal{F}^{2}(m_{K^{*}},q^2).
\end{eqnarray}}}
In the above expressions, $\mathcal{F}(m_{i},q^2)$ etc denotes the
form factors which compensate the off-shell effects of mesons at the
vertices and are written as \cite{HY-Chen,FF}
\begin{eqnarray}
\mathcal{F}(m_{i},q^2)=\bigg(\frac{\Lambda^{2}-m_{i}^2
}{\Lambda^{2}-q^{2}}\bigg)^{n},
\end{eqnarray}
where $\Lambda$ is a phenomenological parameter. As $q^2\to 0$ the
form factor becomes a number. If $\Lambda\gg m_{i}$, it becomes
unity. As $q^2\rightarrow\infty$, the form factor approaches to
zero. As the distance becomes very small, the inner structure would
manifest itself and the whole picture of hadron interaction is no
longer valid. Hence the form factor vanishes and plays a role to cut
off the end effect. The expression of $\Lambda$ is \cite{HY-Chen}
\begin{eqnarray}
\Lambda(m_{i})=m_{i}+\alpha \Lambda_{QCD},\label{parameter}
\end{eqnarray}
where $m_{i}$ denotes the mass of exchanged meson and $\alpha$ is a
phenomenological parameter. Although we use $\Lambda_{QCD}=220$ MeV,
the range of $\Lambda_{QCD}$ can be taken into account through the
variation of the parameter $\alpha$. In this work, we adopt the
monopole form factor $\mathcal{F}(m_{i},q^2)={(\Lambda^{2}-m_{i}^2)
}/{(\Lambda^{2}-q^{2})}$, where $\alpha$ is of order unity and its
range is around $0.8<\alpha<2.2$ \cite{HY-Chen}.

\subsection{Dispersive contribution}

As the bridge between the dispersive part of FSI amplitude and the
absorptive part, the dispersion relation is
\begin{eqnarray}
\mathbf{Dis}
\mathcal{M}(m_X)=\frac{1}{\pi}\int^{\infty}_{s_{0}}\frac{\mathbf{Abs}
\mathcal{M}(s)}{s-m_{X}^2}ds,
\end{eqnarray}
with
\begin{eqnarray}
\mathbf{Abs}
\mathcal{M}(s)&=&\Big\{\mathbf{Abs}^{(a)}[\rho^{+}\rho^{-},K^{0}]
\nonumber\\&&+\mathbf{Abs}^{(b)}[\rho^{+}\rho^{-},K^{*0}]\Big\}\exp(-\beta
|\mathbf{k}|^2),\nonumber
\end{eqnarray}
where $\mathbf{k}=\sqrt{{s}/{4}-m_{\rho}^{2}}$ is the three
momentum of $\rho^{\pm}$ in the rest frame of $\rho(1450,1700)$.
The exponential reflects the dependence of the interaction between
$\rho(1450,1700)$ and $\rho^{\pm}$ on $\mathbf{k}$, which also
plays the role of the cutoff. The factor $\beta$ is related to the
radius of interaction $R$ by $\beta = R/6$ \cite{Pennington}.

\begin{figure}[htb]
\begin{center}
\begin{tabular}{c}
\scalebox{0.8}{\includegraphics{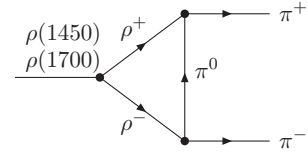}}\end{tabular}
\end{center}
\caption{The decay of $\rho(1450,1700)\to \pi^{+}\pi^{-}$ through
the $\rho^{\pm}$ pair.}\label{pipi}
\end{figure}

Using the same formalism, we calculate the decay amplitude of
$\rho(1450,1700)\to\rho^+\rho^-\to \pi^+\pi^-$ as depicted in Fig.
\ref{pipi}.

\section{Results and discussions}\label{sec3}

Using $\Gamma(\phi\to K^{+}K^{-})=2.09$ MeV \cite{PDG}, we obtain
$g_{\phi K^{+}K^{-}}=5.55$. In the limit of SU(3) symmetry, we take
$\sqrt{2}g_{\rho^{0} K^{\pm}K^{\mp}}=g_{\phi K^{\pm}K^{\mp}}$.
$g_{\rho^{\pm}K^{\mp}K^{*0}}=6.48$ GeV$^{-1}$ \cite{zhao}.
$g_{\rho(1450)\rho^+\rho^-}=1.53$ and
$g_{\rho^{+}\pi^{0}\pi^{+}}=g_{\rho^{-}\pi^{0}\pi^{-}}=11.5$
\cite{PDG}.

In Fig. \ref{diagrams}, we show the dependence of the width of
$\rho(1450, 1700)\to \rho^+\rho^- \to K^+ K^-$ on the mass of
$\rho(1450, 1700)$ with the typical parameters $\alpha=1.0,1.5, 2.0$
and $\beta=0.2, 0.4, 0.8$ GeV$^{-2}$ \cite{Pennington}.
\ref{diagrams-1} is the dependence of the width of
$\rho(1450,1700)\to\rho^+\rho^-\to \pi^+\pi^-$ on the mass of
$\rho(1450, 1700)$ with several typical values.
\begin{widetext}
\begin{center}
\begin{figure}[htb]
\begin{tabular}{cccccccc}
\scalebox{0.55}{\includegraphics{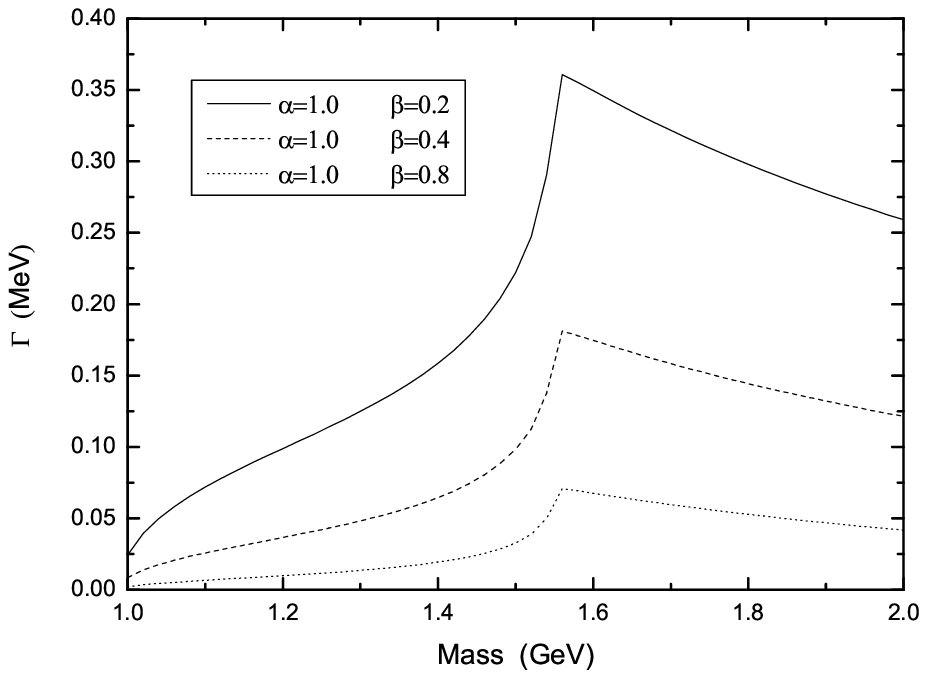}}&\scalebox{0.55}{\includegraphics{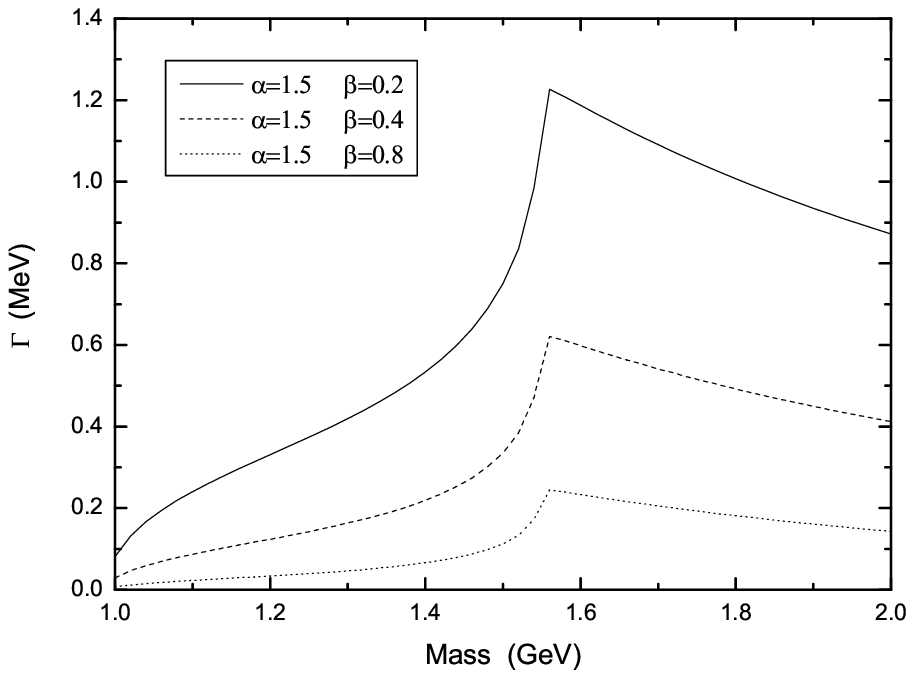}}&\scalebox{0.55}{\includegraphics{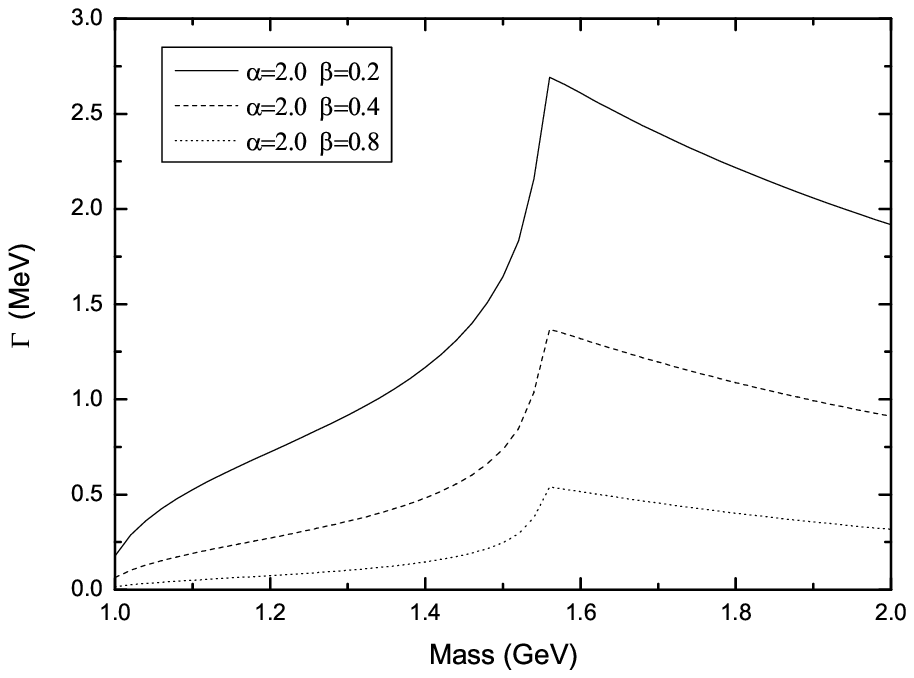}}
\end{tabular}
\caption{The dependence of the width of $\rho(1450, 1700) \to K^+
K^-$ on the the mass of $\rho(1450, 1700)$ with several typical
values of $\alpha$ and $\beta$.\label{diagrams}}
\end{figure}
\begin{figure}[htb]
\begin{tabular}{c}
\scalebox{1}{\includegraphics{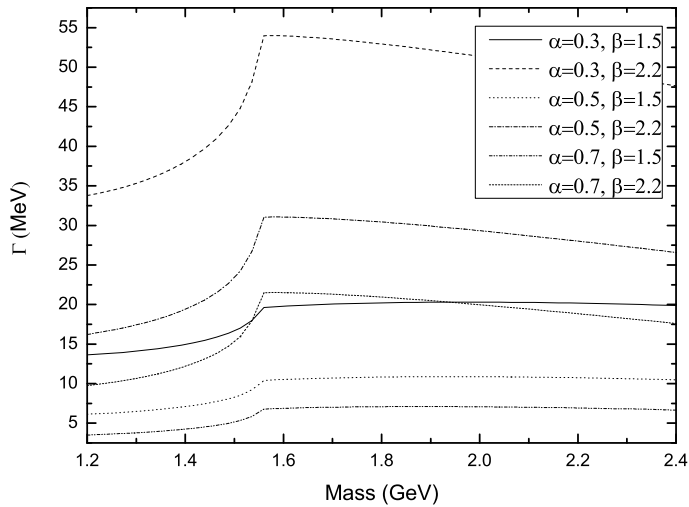}}
\end{tabular}
\caption{The dependence of the width of $\rho(1450, 1700) \to \pi^+
\pi^-$ on the the mass of $\rho(1450, 1700)$ with several typical
values of $\alpha$ and $\beta$.\label{diagrams-1}}
\end{figure}
\end{center}
\end{widetext}

In this short note, we revisit the final state interaction of the
$\rho(1450,1700)\to \rho^+\rho^-$ decay. Different from our former
work \cite{Liu}, we consider the additional contribution from the
dispersive part. Our result shows that there exists an enhancement
around 1580 MeV as shown in Fig. \ref{diagrams}. Such an enhancement
occurs with the opening of the $\rho\rho$ channel.

The decay width of $\rho(1450,1700)\to K^+ K^-$ from the FSI
effect is a few MeV only. If the width of $\rho(1450,1700)$ is 300
MeV, the branching ratio of $\rho(1450,1700)\to A+B\to K^{+}K^{-}$
is about $10^{-2}$. With $B[J/\psi\to \pi+\rho(1450,1700)]$
roughly around $10^{-3}$ \cite{PDG}, $B[J/\psi\to
\pi^{0}+\rho(1450,1700)]\cdot B[\rho(1450,1700)\to AB\to K^{+}
{K}^{-}]$ is about $10^{-5}$. The dispersive contribution enhances
the branching ratio of $\rho(1450,1700)\to K^+K^-$ by two orders
than that in Ref. \cite{Liu}. However, such a ratio is still far
less than experimental value $B[J/\psi\to \pi+X(1576)]\cdot
B[X(1576)\to K^{+}{K}^{-}]=(8.5\pm 0.6^{+2.7}_{-3.6})\times
10^{-4}$, although the $K^+K^-$ spectrum from the FSI effect of
$\rho(1450,1700)$ decays mimics the observed broad spectrum from
BES's measurement.

Throughout our calculation, we ignored the direct coupling between
$\rho(1450, 1700)$ and $K\bar K$. Recently, Li argued that
$\rho(1450, 1700)$ can have strong coupling with $K^{+}K^{-}$ at the
tree level \cite{Li}. Adding this contribution certainly increases
the branching ratio. However, the experimental upper limit of
$B[\rho(1450)\to K\bar K]$ is $1.6\times 10^{-3}$ and $K\bar K$ is
not one of the dominant decay modes of $\rho(1700)$ \cite{PDG}.
Clearly, future BESIII high-statistics data around 1.6 GeV in the
$K\bar K$ channel will be very helpful in the clarification of
$X(1576)$. We also calculate the decay amplitude of
$\rho(1450,1700)\to\rho^+\rho^-\to \pi^+\pi^-$ using the same
technique. There exists one similar enhancement around 1580 MeV,
which is shown in Fig. \ref{diagrams-1}. This enhancement will be
useful to test if X(1576) arises from the FSI effect.

\vfill

\section*{Acknowledgments}

We thank Prof. B.S. Zou for the useful suggestion. This project
was supported by the National Natural Science Foundation of China
under Grants 10421503 and 10625521, and the China Postdoctoral
Science foundation (20060400376).

\end{document}